%% file: main.tex
\documentclass{article}

\PassOptionsToPackage{numbers,compress}{natbib}

\usepackage[preprint]{neurips_2026}
\usepackage[utf8]{inputenc}
\usepackage[T1]{fontenc}

\usepackage{url}
\usepackage{booktabs}
\usepackage{amsfonts}
\IfFileExists{nicefrac.sty}{\usepackage{nicefrac}}{}
\usepackage{microtype}
\usepackage[table]{xcolor}
\usepackage{multirow}
\usepackage{array}
\usepackage{tabularx}
\usepackage{longtable}
\usepackage{caption}
\usepackage{calc}
\usepackage{changepage}
\usepackage{float}
\usepackage{tikz}
\usetikzlibrary{arrows.meta,positioning,shapes.geometric,fit,calc}
\usepackage{hyperref}
\hypersetup{
  hidelinks,
  pdftitle={Who Said What, and Will It Be Remembered? Evaluating Persistent Speaker Attribution Across Meetings},
  pdfauthor={Shantanu Vispute; Aditya M. Mishra; Siddhartha Saxena},
  pdfsubject={Technical report on persistent speaker attribution for ambient meeting transcripts},
  pdfkeywords={speaker attribution, speaker diarization, ambient meeting transcription, persistent identity, SI-cpWER, ThyVoice}
}

\providecommand{\tightlist}{%
  \setlength{\itemsep}{0pt}\setlength{\parskip}{0pt}%
}

\newcommand{\tableinfo}[1]{%
  \par\vspace{3pt}%
  \begin{minipage}{\linewidth}
  \footnotesize\textit{Notes.} #1
  \end{minipage}%
  \par\vspace{4pt}%
}

\newenvironment{widetableblock}{%
  \begin{adjustwidth}{-0.5in}{-0.5in}%
  \setlength{\hsize}{\linewidth}%
  \LTleft=-0.5in plus 1fill\relax
  \LTright=0.5in plus 1fill\relax
}{%
  \end{adjustwidth}%
}

\definecolor{thineblue}{RGB}{49,92,172}
\definecolor{thinegreen}{RGB}{44,137,90}
\definecolor{thineamber}{RGB}{178,118,42}
\definecolor{thineline}{RGB}{105,116,135}

\title{Who Said What, and Will It Be Remembered? Evaluating Persistent Speaker Attribution Across Meetings}

\author{%
  Shantanu Vispute\thanks{Equal contribution.}\\
  \textit{Foyer}\\
  \texttt{shantanu@foyer.work}
  \And
  Aditya M. Mishra\footnotemark[1]\thanks{Work performed while at Foyer.}\\
  \textit{Foyer}\\
  \texttt{aditya.mishra@foyer.work}
  \And
  Siddhartha Saxena\\
  \textit{Foyer}\\
  \texttt{siddhartha@foyer.work}
}

\begin{document}

\maketitle

\begin{abstract}
\input{sections/abstract}
\end{abstract}

\section{Introduction}
\input{sections/introduction}

\section{Related Work}
\input{sections/related_work}

\section{Persistent-Attribution Evaluation}
\input{sections/dataset}

\section{ThyVoice Reference System}
\input{sections/methodology}

\section{Results and Layered Diagnostics}
\input{sections/experiments_results}

\section{Discussion}
\input{sections/discussion}

\section{Limitations}
\input{sections/limitations}

\section{Conclusion}
\input{sections/conclusion_future_work}

\section*{Acknowledgments}
AI writing assistants were used for copy-editing; all system design, experiments, analysis, and results are the authors' own.

\bibliographystyle{plainnat}
\setlength{\bibsep}{4pt}
\bibliography{references}

\clearpage
\appendix
\input{sections/appendix}

\end{document}

%% file: sections/abstract.tex
Speech transcripts used as long-term memory must preserve both words and stable speaker identities. Existing meeting-transcription metrics either ignore speakers or remap anonymous speakers independently in each recording, so they cannot measure whether the same person retains one identity across meetings. We evaluate persistent speaker attribution with \textbf{Speaker Identified cpWER} (SI-cpWER), which scores a corpus under one global speaker-ID assignment. The benchmark covers five commercial diarize-then-identify cascades, two open academic baselines, and ThyVoice on the full 129-meeting CHiME-8 NOTSOFAR evaluation set in clean and noise-augmented form, plus CHiME-6. ThyVoice is our end-to-end reference system; it repairs overlap and gates the evidence used to create and update voiceprints. Requiring persistent identity changes the commercial ranking: ThyVoice records lower SI-cpWER than every evaluated commercial cascade in all three conditions and the lowest mean in the full panel, 47.13 versus 54.75 for the next system. Complementary lexical, diarization, per-recording attribution, and speaker-clustering diagnostics characterize upstream error surfaces in the final attributed record. These results show why persistent attribution must be evaluated directly in systems that reuse conversations across time.

%% file: sections/introduction.tex
Automatic speech recognition (ASR) benchmarks typically emphasize word-level correctness. This is necessary but insufficient for systems that turn conversations into durable, queryable records. An ambient assistant that answers \emph{who committed to what} or \emph{what this person decided last week} needs a faithful record of both the words and the person who spoke them.

This paper studies the transcript layer of that record. Reusable attribution requires the same person to retain one identity across sessions; otherwise, facts cannot be linked reliably across meetings. The distinction also changes the consequence of errors: a missing word reduces recall, while a misattributed word can attach a fact, preference, or commitment to the wrong person.

We ask two evaluation questions. First, how does requiring one persistent identity map change system rankings relative to per-recording attribution? Second, which lexical, local-speaker, and identity failures explain the persistent-attribution result? Our contributions are:

\begin{itemize}
\tightlist
\item
  \textbf{A persistent-attribution evaluation.} We adopt \textbf{Speaker Identified cpWER} (SI-cpWER) --- cpWER under one corpus-global speaker-ID map --- as the primary metric for reusable speaker attribution.
\item
  \textbf{A layered benchmark.} We evaluate five commercial diarize-then-identify cascades and two open academic baselines on CHiME-8 NOTSOFAR in clean and fixed noise-augmented conditions, plus CHiME-6. SI-cpWER, cpWER, DER/JER, WER, and full-corpus speaker-clustering and overlap diagnostics characterize complementary error surfaces in the final persistent record.
\item
  \textbf{ThyVoice as a reference system.} ThyVoice uses overlap-repaired evidence and gated identity updates; it leads commercial SI-cpWER in all three conditions and the full-panel mean.
\end{itemize}

We evaluate systems end-to-end on single-channel meeting audio, where diarization and speaker identity must be inferred from one ambient stream. We first compare persistent and per-recording attribution, then use local-speaker and lexical diagnostics to interpret the result.

%% file: sections/related_work.tex
\textbf{Speaker-attributed and multi-talker ASR.} Systems that answer \emph{who said what} combine recognition with speaker assignment in three broad ways. Reconcile-by-timestamp cascades transcribe and diarize independently, then align words to speaker turns, as in WhisperX and PyannoteAI's STT orchestration \cite{whisperx}. Segment-first systems diarize before recognition and transcribe per-speaker audio or target-speaker masks; Diarization-Conditioned Whisper and its speaker-enhanced SE-DiCoW variant are examples \cite{dicow}. A third family folds attribution into the recognizer through multiple streams, speaker tokens, or structured generation, including permutation-invariant ASR, serialized-output training, Sortformer, Granite Speech, and VibeVoice-ASR \cite{pitasr, sot, tsot, sortformer, granite, granitesaa, vibevoiceasr}. These arrangements expose different failure boundaries: diarizing first can truncate words at segment boundaries, while diarizing afterwards must reconcile independently produced word and speaker timelines \cite{diarizationsurvey}.

\textbf{Overlap-aware diarization and enrollment safety.} Neural diarization has moved from clustering-only pipelines toward frame-wise multi-speaker activity prediction, including EEND and pyannote's powerset segmentation \cite{eend, pyannote21, powersetdiarization}. Our evaluated pyannote baseline uses the proprietary Precision-2 API rather than the OSS pipeline \cite{precision2}. Overlap remains a central stress case because timestamp cascades must reconcile simultaneous words and turns, target-speaker recognizers depend on diarization masks, and serialized-output systems must linearize concurrent speech. VibeVoice-ASR reports that its serialized stream does not explicitly handle overlap, while IBM's speaker-attributed ASR setup discards, serializes, or filters overlap during data construction \cite{vibevoiceasr, granitesaa}. The same boundary matters for persistent identity: mixed speech presented as single-speaker evidence cannot be excluded before enrollment. PyannoteAI therefore recommends voiceprint enrollment audio containing only the target speaker \cite{pyannotevoiceprint}.

\textbf{Persistent identity and modularity.} Within-recording speaker labels do not by themselves preserve a person across recordings. Classical profile-based systems assume a known roster, while ambient capture must discover speakers online and retain stable identities as they recur \cite{diarizationsurvey}. False splits fragment one person's history; false merges attach evidence from different people to one identity. A standard implementation combines pretrained speaker embeddings with similarity-based enrollment and matching, but its reliability still depends on diarization, evidence selection, thresholds, and update policy \cite{ecapa, xvector, ge2e, speechbrain, wespeaker}. Component access also differs: open cascades such as pyannote.audio and WhisperX expose replaceable stages, while hosted diarized-STT and end-to-end systems usually expose a fixed pipeline \cite{whisperx, pyannoteaudio}.

Together, these lines leave a practical evaluation gap. Per-recording metrics can remap anonymous speakers independently, while profile-based systems can assume a fixed enrolled roster. Persistent ambient capture instead requires an identity namespace that is discovered online and remains stable across recordings. Our evaluation preserves each system's exposed surface while separating lexical, local-speaker, per-recording, and persistent-attribution evidence.

%% file: sections/dataset.tex
\subsection{Persistent-attribution task and headline metrics}\label{persistent-attribution-task-and-headline-metrics}

Standard WER measures lexical correctness without speaker identity. Concatenated minimum-permutation word error rate (cpWER) concatenates each speaker's words within a scored sample and chooses the one-to-one speaker assignment that minimizes word error for that sample. Other speaker-attributed metrics include SA-WER and tcpWER; tcpWER additionally constrains matches by time \cite{kanda20saasr, chime6, meeteval, notsofar, chime8dasr}. These metrics do not test whether the same identity is reused across recordings.

Our primary metric is \textbf{Speaker Identified cpWER (SI-cpWER)}: cpWER computed under one \textbf{global speaker-ID map} for the corpus. Headline cpWER scores each sample (a NOTSOFAR meeting or CHiME-6 chunk) using the system's pre-identification \texttt{local\_speaker} labels and chooses a fresh hypothesis-to-reference assignment for that sample. SI-cpWER instead scores the post-identification persistent \texttt{speaker} labels under one corpus-global, one-to-one assignment. It therefore measures both the attributed words and whether each person retains one identity, without requiring a pre-enrolled speaker roster.

Let \(N\) be the number of reference words and let \(S,D,I\) be substitutions, deletions, and insertions after speaker alignment. Headline cpWER is \[
\mathrm{cpWER} = \frac{S_{\mathrm{local}} + D_{\mathrm{local}} + I_{\mathrm{local}}}{N}.
\] SI-cpWER uses one identity map for the full corpus: \[
\mathrm{SI\text{-}cpWER} = \frac{S_{\mathrm{global}} + D_{\mathrm{global}} + I_{\mathrm{global}}}{N}.
\]

For each hypothesis-to-reference speaker pair, we compute word-edit cost within each scored sample and sum those costs across samples. A Hungarian solver selects the minimum-cost global one-to-one assignment; sample boundaries remain intact, so transcripts are not concatenated across meetings or chunks. Dummy assignments handle unequal speaker inventories: unmatched reference speakers contribute deletions, and unmatched hypothesis speakers contribute insertions. Explicit unknown hypothesis labels remain in the headline score.

Each condition begins with an empty identity namespace and processes samples in one fixed order, modeling recordings as they arrive over time without pre-enrolled speakers. The first accepted enrollment initializes a voiceprint from eligible audio; its quality can affect subsequent matches, so changing the order can change later assignments and SI-cpWER.

The difference SI-cpWER \(-\) cpWER summarizes the net score change between the pre-identification and post-identification surfaces. A positive value indicates added error; a value near zero indicates little net change; and a small negative value can occur when identity resolution merges local fragments before global scoring.

\subsection{Datasets and conditions}\label{datasets-and-conditions}

We use the held-out CHiME-8 NOTSOFAR evaluation set from the official eval-with-ground-truth release \cite{notsofar, chime8dasr}. It matches ambient capture because it provides natural office meetings with overlapping conversation, room-device audio, and speaker-attributed references. Its participant aliases persist across meetings, so one reference speaker ID denotes the same person throughout the scored corpus, as required for global identity scoring. NOTSOFAR provides utterance-level timing; we interpolate word-level reference times uniformly within each utterance, using the same convention as for CHiME-6. Approximately 30\% of its evaluation speech is overlapped \cite{chime8dasr}.

Headline results and the persistent identity-count diagnostic use all 129 meetings. Ablations use the first 20 meetings in native order, without shuffling or selection by difficulty. Their recurring five-speaker cast makes local speaker counts and global identity fragmentation interpretable. These ablations provide directional evidence on this subset; they do not establish the size of the effect over the full corpus.

We construct two evaluation conditions. For \textbf{Clean}, we select one single-channel, non-close-talk device recording per meeting under a fixed random seed. This avoids evaluating several recordings of the same meeting while preserving the original meeting audio and reference transcript. We intentionally do not use multi-channel beamforming, array geometry, or device fusion; every system receives the same single recording stream.

Real ambient capture is rarely as controlled as the Clean condition. A deployed recorder may capture steady room noise, nearby non-target speech, and short acoustic events from the environment. We therefore construct \textbf{Noisy} by applying a fixed augmentation to the Clean audio while keeping the transcript, word timestamps, and speaker labels unchanged. The augmentation stresses different failure modes: WHAM ambient noise adds a continuous environmental noise bed \cite{wham}; MUSAN speech babble adds competing non-target speech \cite{musan}; and MUSAN transient events add brief local disturbances such as keyboard activity, desk contact, chair movement, and door sounds. The noise pools are the WHAM! validation split (4,444 clips), MUSAN \texttt{speech} (426 clips) for babble, and MUSAN \texttt{noise} (930 clips) for transients; MUSAN music is not used. Every random choice --- device selection, noise clips and offsets, per-meeting SNR, and transient onsets --- was fixed by a seeded procedure for the reported runs.

Let \(x[t]\) be the Clean waveform. We construct a composite noise signal \[
n[t] = \sqrt{0.70}\,w[t] + \sqrt{0.15}\,b[t] + \sum_j \sqrt{0.15}\,r_j[t-\tau_j],
\] where \(w[t]\) is WHAM ambient noise, \(b[t]\) is MUSAN speech babble, and \(r_j[t-\tau_j]\) are short MUSAN transient events. Each component is RMS-normalized to unit energy before weighting, so the squared weights are the intended energy shares. The babble component is additionally low-pass filtered at 2 kHz and reverberated with \(\mathrm{RT}_{60}=0.4\,\mathrm{s}\) so that the competing speech is muffled and spatially indirect rather than clean foreground speech. Transient onsets \(\{\tau_j\}\) are drawn from a Poisson process at 4 events per minute. The noisy waveform is \[
y[t] = x[t] + \alpha\, n[t],
\] where \(\alpha\) is chosen to match the target SNR, sampled per meeting from \(\mathcal{N}(10,3^2)\) and clipped to \([3,18]\) dB (realized mean 10.1 dB, range 3.3--17.5). Because Clean and Noisy share the same reference transcript and speaker labels, differences between the two conditions isolate robustness to acoustic degradation rather than changes in annotation.

\begin{table}[H]
\centering
\begingroup\small
\setlength{\tabcolsep}{12pt}
\caption{Evaluation-set statistics for the NOTSOFAR conditions.}\label{tab:dataset}
\begin{tabular}{@{}ll@{}}
\toprule
Statistic & Value \\
\midrule
Meetings (total / evaluated) & 129 / 129 \\
Speakers (disjoint test) & 12 \\
Speakers per meeting & 4.65 (range 3--7) \\
Audio duration & 13.34 h \\
Scored reference words & 188,036 (218,825 before scoring normalization) \\
Overlapped speech & $\approx$30\% (NOTSOFAR eval) \\
Noisy SNR & mean 10.1 dB, range {[}3.3, 17.5{]} \\
Ablation subset & first 20 meetings; 5 recurring speakers; 29,714 words \\
\bottomrule
\end{tabular}
\endgroup
\end{table}

The second corpus and third evaluation condition is CHiME-6 \cite{chime6}: real dinner-party conversations recorded in homes on far-field devices, with four participants per session and heavily overlapped, unstaged speech (roughly a third of speech time in the sessions we use). Where the Noisy condition is a controlled augmentation of office-meeting audio, CHiME-6 is true far-field capture, and it differs sharply from NOTSOFAR in acoustics, room conditions, and conversational style. We use the two evaluation sessions (S01 and S21; 56,886 reference words) as single-channel audio. Because each session is a single recording longer than two hours rather than a sequence of meetings, we simulate the longitudinal setting explicitly: every system processes the session as a sequence of ${\sim}300$\,s chunks fed serially against a shared voiceprint bank, so persistent identity must survive chunk boundaries the way it must survive meeting boundaries on NOTSOFAR. The two sessions have disjoint speaker sets, but the voiceprint bank is shared across both sessions as well --- deliberately, for increased difficulty: the identity layer must keep each session's speakers consistent without falsely matching them to the other session's enrolled strangers. For the headline cpWER and SI-cpWER results, cut points target 300 s and snap to the nearest reference word boundary within a $\pm$10 s tolerance, so no reference word is split across chunks. Comparison cascades use stitched-30s enrollment; ThyVoice uses its native pooled multi-vector enrollment. The dedicated WER and DER/JER diagnostics use the saved full-session or chunked component outputs available for each evaluated system; their scored surfaces are identified in the corresponding tables.

Two scoring notes. First, the official CHiME-6 utterance annotations carry no word-level timestamps. To score chunked hypotheses against time-correct references, we derive word timings from the official utterance annotations using meeteval's equidistant-intervals pseudo-word-timing convention \cite{meeteval} --- words spread uniformly within each annotated segment, a convention shown to score on par with exact word timings and used in official CHiME ranking metrics --- and chunk cuts snap to word boundaries. Second, absolute error rates on CHiME-6 are high for every system: the official challenge's unsegmented-track baseline scored roughly 78\% cpWER on the evaluation set \cite{chime6}, and that baseline had access to multichannel array audio with dereverberation and beamforming, whereas every system here receives one raw channel. The panel should therefore be read comparatively, not against clean-meeting expectations.

\subsection{Systems and identity protocol}\label{systems-and-identity-protocol}

We compare ThyVoice against five commercial diarize-then-identify cascades and two open academic baselines. In each comparison cascade, a provider produces a diarized transcript (words tagged with local speaker clusters) for each meeting, and a shared PyannoteAI \textbf{voiceprint} layer then resolves those local clusters to persistent identities. The voiceprint bank is shared across all meetings and the stacks run serially, so identity must persist across recordings, not just within one. For each local cluster the layer queries pyannote \texttt{/identify} using \texttt{matching.threshold=50} on pyannote's 0--100 confidence scale. If no voiceprint clears the threshold and the cluster has sufficient eligible audio, the layer enrolls a new voiceprint (\texttt{/voiceprint}) from up to 30 s of that speaker's audio; otherwise the cluster remains unknown. The full-scale runs stitch (concatenate) eligible speaker segments up to 30 s (\textbf{stitched-30s} enrollment), while the single-longest-turn variant is kept as an ablation arm. For the pyannote stack, eligible enrollment evidence is derived from its overlap-aware standard diarization: every interval in which another diarized speaker is simultaneously active is removed before stitching. Words inherit their cluster's resolved identity. In total we evaluate \textbf{five commercial diarize-then-identify cascades} --- ElevenLabs, PyannoteAI, AssemblyAI, Deepgram, and OpenAI --- all sharing the same PyannoteAI identity backend, alongside ThyVoice and \textbf{two open academic baselines} (SE-DiCoW and WhisperX) run through the same shared identity backend. All systems are scored under the same metrics and reported together in the results.

When the voiceprint bank exceeds pyannote \texttt{/identify}'s 50-entry request limit, the accepted full-corpus runs search the bank in batches and select the best above-threshold match before enrolling a new identity. Appendix A records the batching rule and provider-specific run details.

\textbf{ElevenLabs + pyannote voiceprints (\texttt{elevenlabs}).} ElevenLabs \textbf{Scribe v2} produces the diarized transcript --- word timestamps and local speaker labels --- and same-speaker words are merged into clusters across gaps up to one second. We keep ElevenLabs' diarization and use pyannote only for identity. The evaluated Scribe output provides one speaker label per word but no separate simultaneous-speaker activity surface, so enrollment uses the emitted cluster intervals directly. Sending the whole meeting to \texttt{/identify} would impose pyannote's own diarization, whose segment boundaries and clusters need not align with ElevenLabs' words; instead we query \texttt{/identify} once per ElevenLabs cluster, using the concatenated audio of that cluster's segments, and attach the returned voiceprint label. This preserves ElevenLabs' word-to-speaker assignment while resolving each cluster to a persistent cross-meeting identity. Because clusters are matched independently, two clusters can resolve to the same identity (merging a speaker ElevenLabs over-split).

\textbf{pyannote STT + pyannote voiceprints (\texttt{pyannote}).} PyannoteAI \textbf{Precision-2} produces diarization and transcription in a single \texttt{/diarize} call with \texttt{transcription=true}, using its default \textbf{Parakeet TDT 0.6B v3} recognizer. Its standard \texttt{diarization} surface can mark simultaneous speakers, while requesting \texttt{exclusive=true} additionally returns an \texttt{exclusiveDiarization} surface with only one active speaker at a time \cite{pyannotestt}. Transcript scoring consumes the emitted \texttt{wordLevelTranscription}. The identity path uses the standard overlap-aware diarization to define local speaker clusters and derives overlap-filtered enrollment evidence by subtracting every interval in which another diarized speaker is active before applying stitched-30s enrollment. Identity is resolved by a full-audio \texttt{/identify} call with \texttt{matching.threshold=50} and \texttt{matching.exclusive=true}; the returned labeled segments are mapped back to the standard diarization clusters by temporal overlap. This is the pyannote-specific orchestration we score, and it differs from the ElevenLabs cascade, where pyannote is used only after ElevenLabs has already produced word-level speaker labels.

\textbf{Other hosted diarized-transcription providers.} The remaining hosted cascades use AssemblyAI \textbf{Universal-3 Pro}, Deepgram \textbf{Nova-3}, and OpenAI \textbf{gpt-4o-transcribe-diarize}. On the evaluated output surfaces, these providers assign one speaker to each emitted word or segment and do not expose a separate simultaneous-speaker activity timeline from which overlap can be removed before enrollment. Their emitted speaker intervals therefore supply the stitched-30s enrollment evidence, which is resolved through the same shared voiceprint backend. Appendix A records the time-specific Deepgram diarizer resolution.

\textbf{ThyVoice (ours).} ThyVoice is evaluated end-to-end on raw meeting audio under the same global speaker-ID scoring as the other systems. It infers speaker activity, builds clean per-speaker audio, transcribes, and resolves each local speaker against its persistent voiceprint store; the next section describes the full pipeline. ThyVoice uses Qwen3-ASR as the recognizer, with Qwen forced alignment to recover word timestamps; this configuration is held fixed across all conditions.

\textbf{Academic baselines.} \textbf{SE-DiCoW} uses the released \texttt{BUT-FIT/SE-DiCoW} checkpoint with its DiariZen md diarization front end. It is BUT-FIT's speaker-enhanced Diarization-Conditioned Whisper \cite{dicow}: the recognizer transcribes each target speaker from speaker-activity masks, a design intended to handle overlap better than timestamp-reconciliation cascades. We use SE-DiCoW as the family's representative open-weights checkpoint: it is comparable to DiCoW v3.3 on NOTSOFAR single-channel on the official model cards, and the family's serialized-output variants (SA-/SOT-DiCoW) have no released checkpoint and report with oracle diarization, where their own results show the DiCoW variants ahead on real meetings \cite{sadicow}. Training mixtures for both the SE-DiCoW recognizer and the evaluated DiariZen checkpoints include NOTSOFAR-1 \cite{sedicow,diarizen}; SE-DiCoW's published NOTSOFAR numbers use tcpWER with a 5-second collar on a challenge subset \cite{sedicow}, a different metric, subset, and protocol from the corpus-global SI-cpWER scoring here. \textbf{WhisperX} is the common open reconcile-by-timestamp reference point: Whisper \texttt{large-v2} transcription, forced alignment, and pyannote Community-1 diarization \cite{whisperx}. Both are resolved to persistent identities by the same shared pyannote voiceprint backend as the commercial cascades, so only their diarization and recognition differ.

\subsection{Supporting diagnostics and repeatability}\label{supporting-diagnostics-and-repeatability}

The supporting metrics diagnose the evidence available before persistent identity resolution. \textbf{IncorrectAssertionRate (IAR)} separates cpWER omissions from asserted errors: \[
\mathrm{IAR} = \frac{S_{\mathrm{local}} + I_{\mathrm{local}}}{N}.
\] Because cpWER is speaker-attributed, substitutions and insertions can reflect lexical errors or real words assigned to the wrong speaker. IAR does not measure global identity drift, which remains part of SI-cpWER.

At the lexical layer, \textbf{WER} scores emitted words without speaker labels. For hosted speaker-labelled outputs, it scores the returned diarized-transcription text; for standalone recognizers, it is recognizer-only WER. At the local-speaker layer, \textbf{DER} measures missed speech, false alarm, and speaker confusion relative to reference speaker time. \textbf{JER} averages one minus intersection-over-union over optimally mapped reference and hypothesis speakers. The primary DER/JER protocol uses zero collar and includes overlap. Native diarization timelines are scored unchanged; one-speaker-per-word outputs are reconstructed into turns with a fixed 0.25 s maximum gap.

The same interval partition supports speaker-count error, fragmentation, merging, cluster purity, single-speaker coverage, and overlap-detection diagnostics. These are within-recording measures. They indicate what local evidence is available before enrollment but do not replace SI-cpWER. Full-corpus overlap detection is reported only for native timelines that can represent simultaneous speakers; overlap is not observable from an exposed one-speaker-per-word surface.

These metrics describe connected evidence surfaces, but system architectures do not execute them in one universal order. Missing or incorrect words constrain the attributed transcript; local speaker merges, confusion, and hidden overlap constrain identity evidence; persistent resolution can add error or repair local fragmentation. Values are compared within each layer and are not added across metrics because their denominators and alignments differ. In main result tables, \textbf{Mean} is the arithmetic average of the Clean, Noisy, and CHiME-6 condition-level rates.

Run-to-run variation matters because hosted recognition, diarization, and voiceprint calls can change across otherwise identical requests, while online enroll-or-match decisions can amplify small upstream changes. We therefore ran three full 129-meeting ElevenLabs draws for each NOTSOFAR condition, where close headline margins warranted repeats. The main results report the corresponding NOTSOFAR means and the single CHiME-6 run. Appendix A records every draw and the exact repeatability scope.

%% file: sections/methodology.tex
\subsection{Gated speaker evidence}\label{gated-speaker-evidence}

We use \textbf{reference system} to mean our end-to-end implementation used to demonstrate the persistent-attribution evaluation, not a standard baseline. ThyVoice is a diarization-guided, segment-first system designed to produce transcripts whose words carry globally stable speaker identities. Its central principle is \textbf{gated speaker evidence}: detect or repair overlap before enrollment, reject ambiguous evidence, and build or update voiceprints only from eligible speaker evidence. Rather than transcribe the mixed audio and label it afterwards, it follows a diarization-before-separation recipe consistent with recent meeting systems \cite{chime8dasr}: diarization marks speaker activity and bounded overlaps, separation runs only where overlap is detected, and ASR consumes clean or confidently recovered per-speaker audio. This section describes the default system used for the headline comparison, with component configurations and gating thresholds held fixed across conditions.

\begin{figure}[H]
\centering
\makebox[\textwidth][c]{\includegraphics[width=6.0in]{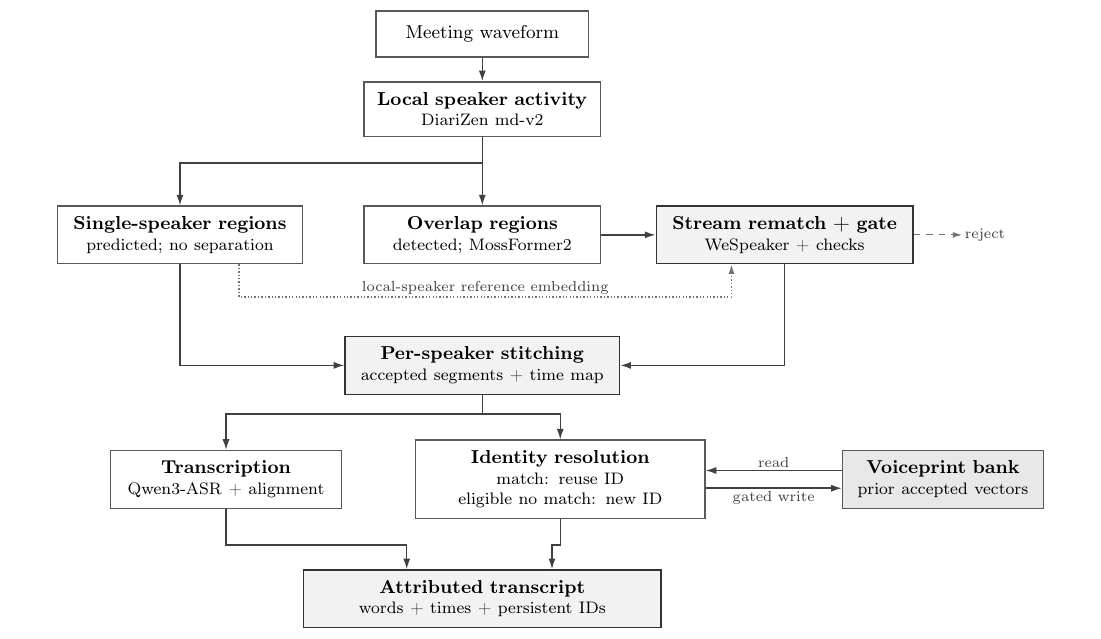}}
\caption{ThyVoice's gated speaker-evidence pipeline.}
\label{fig:thyvoice-pipeline}
\end{figure}

\textbf{Pipeline components.} The DiariZen md-v2 diarizer \cite{diarizen} supplies overlap-aware speaker activity and temporary per-meeting labels. MossFormer2 \cite{mossformer2} separates bounded two-speaker regions. A WeSpeaker SimAM-ResNet100 speaker embedder (the \texttt{voxblink2\_samresnet100\_ft} checkpoint) \cite{wespeaker} supports separator-stream rematching and persistent identity. Qwen3-ASR \cite{qwen3asr} recognizes stitched per-speaker audio. Each stage exposes evidence used by the next stage. Figure~\ref{fig:thyvoice-pipeline} summarizes the pipeline.

\vspace{-12pt}
\enlargethispage{\baselineskip}

\subsection{Overlap repair and stream rematching}\label{overlap-repair-and-stream-rematching}

Overlap is unsafe identity evidence until it is repaired. Each region is classified as single-speaker or overlapping. Clean intervals are used directly, but raw overlap is excluded from enrollment. For bounded two-speaker overlaps, ThyVoice applies source separation. Because separator stream order is arbitrary, each separated stream is embedded with SimAM-ResNet100 and cosine-matched to clean local-speaker evidence using a cosine-similarity threshold of 0.25 before it can re-enter that speaker's pool. Silent, duplicated, ambiguous, or poorly matched streams are rejected because single-channel separation can emit duplicate or non-speech content \cite{diarizationsurvey}. ThyVoice therefore recovers overlapped speech only when it can be assigned confidently; otherwise it omits the uncertain evidence.

\subsection{Pooled transcription evidence}\label{pooled-transcription-evidence}

For each local speaker, ThyVoice pools all clean single-speaker segments with only the separated-overlap streams that passed the assignment gates. It concatenates them into one per-speaker stream, inserts silence between pieces, and retains a map from stitched time to the meeting timeline. Pooling provides more phonetic and acoustic coverage than a single turn while avoiding hidden overlap and rejected separator artifacts \cite{diarizationsurvey}. Each stream is transcribed independently with Qwen3-ASR. Forced alignment recovers word times in stitched-stream time, which are then remapped to the meeting timeline.

\subsection{Persistent identity matching and updates}\label{persistent-identity-matching-and-updates}

The pooled evidence is resolved against a persistent \textbf{voiceprint store} by similarity. A match at cosine similarity of at least 0.5 reuses an existing identity; otherwise, eligible evidence can initialize a new identity. Matching and profile updates are separate decisions: adding a matched vector to an existing identity requires similarity of at least 0.7, clean or confidently separated speech, valid speech activity, and non-duplicated streams. Raw overlap and rejected separated channels are never upserted. This gate retains several clean vectors per identity and is intended to reduce identity drift as audio accumulates.

The final output is a normal meeting transcript on the original timeline, with words attached to globally consistent speaker identities.

%% file: sections/experiments_results.tex
\subsection{Persistent attribution}\label{persistent-attribution}

\begingroup\small
\begin{longtable}[]{@{}lrrrr@{}}
\caption{SI-cpWER by condition.}\label{tab:all-systems-si}\tabularnewline
\toprule\noalign{}
System & Clean $\downarrow$ & Noisy $\downarrow$ & CHiME-6 $\downarrow$ & Mean $\downarrow$ \\
\midrule\noalign{}
\endfirsthead
\toprule\noalign{}
System & Clean $\downarrow$ & Noisy $\downarrow$ & CHiME-6 $\downarrow$ & Mean $\downarrow$ \\
\midrule\noalign{}
\endhead
\bottomrule\noalign{}
\endlastfoot
\textbf{ThyVoice} & 34.89 & 51.27 & 55.24 & 47.13 \\
ElevenLabs & 36.13 & 53.89 & 74.22 & 54.75 \\
PyannoteAI & 44.80 & 69.35 & 65.30 & 59.82 \\
AssemblyAI & 48.26 & 58.30 & 77.41 & 61.32 \\
Deepgram & 58.71 & 79.16 & 90.03 & 75.97 \\
OpenAI & 61.09 & 76.01 & 80.47 & 72.52 \\
SE-DiCoW & 32.24 & 49.54 & 83.88 & 55.22 \\
WhisperX & 50.44 & 65.25 & 86.13 & 67.27 \\
\end{longtable}
\endgroup

Table~\ref{tab:all-systems-si} shows that ThyVoice has lower SI-cpWER than every evaluated commercial cascade in all three conditions and the lowest SI-cpWER mean in the full panel. SE-DiCoW leads the full panel on both NOTSOFAR conditions, while ThyVoice leads it on CHiME-6.

\subsection{Diagnosing persistent attribution}\label{diagnosing-persistent-attribution}

We diagnose persistent attribution through per-recording attribution, local speaker activity, and lexical recognition. The following tables score the configured component or provider output available at each layer; these diagnostic surfaces are not additional end-to-end systems. Each upstream layer constrains the final attributed record, but the dedicated component runs do not recreate one complete ThyVoice row.

\subsubsection{Per-recording speaker attribution}\label{per-recording-speaker-attribution}

\begingroup\small
\begin{longtable}[]{@{}lrrrr@{}}
\caption{cpWER by condition.}\label{tab:all-systems-cpwer}\tabularnewline
\toprule\noalign{}
System & Clean $\downarrow$ & Noisy $\downarrow$ & CHiME-6 $\downarrow$ & Mean $\downarrow$ \\
\midrule\noalign{}
\endfirsthead
\toprule\noalign{}
System & Clean $\downarrow$ & Noisy $\downarrow$ & CHiME-6 $\downarrow$ & Mean $\downarrow$ \\
\midrule\noalign{}
\endhead
\bottomrule\noalign{}
\endlastfoot
\textbf{ThyVoice} & 35.53 & 51.42 & 53.76 & 46.90 \\
ElevenLabs & 33.57 & 43.58 & 44.25 & 40.47 \\
PyannoteAI & 39.51 & 52.71 & 56.43 & 49.55 \\
AssemblyAI & 37.24 & 51.36 & 63.77 & 50.79 \\
Deepgram & 55.14 & 74.08 & 88.87 & 72.70 \\
OpenAI & 47.22 & 63.86 & 71.45 & 60.84 \\
SE-DiCoW & 25.62 & 43.12 & 56.88 & 41.87 \\
WhisperX & 50.00 & 60.29 & 67.35 & 59.21 \\
\end{longtable}
\endgroup

Table~\ref{tab:all-systems-cpwer} shows that ElevenLabs has the lowest commercial cpWER in every condition, in contrast to the SI-cpWER ranking in Table~\ref{tab:all-systems-si}. On CHiME-6, its SI-cpWER is approximately 30.0 percentage points above its local cpWER, compared with 1.5 points for ThyVoice. Strong local transcription and attribution therefore do not guarantee persistent identity. SE-DiCoW's NOTSOFAR lead disappears already at the CHiME-6 cpWER stage, before persistent identification adds further error. Different acoustics may contribute, but these runs do not isolate the cause. Identity resolution can add attribution error or repair local over-segmentation, so the cpWER-to-SI-cpWER change is not strictly monotonic. Appendix B reports the full condition-level S/N, D/N, I/N, and IAR values.

\subsubsection{Persistent identity counts}\label{persistent-identity-counts}

\noindent\begin{minipage}{\linewidth}
\begingroup\small
\setlength{\tabcolsep}{6pt}
\captionof{table}{Persistent identities resolved by condition.}\label{tab:persistent-identity-counts}
\noindent\makebox[\linewidth][c]{%
\begin{tabular}{@{}lrrr@{}}
\toprule
System & Clean identities (ref. 12) & Noisy identities (ref. 12) & CHiME-6 identities (ref. 8) \\
\midrule
Reference & 12 & 12 & 8 \\
\textbf{ThyVoice} & 14 & 21 & 17 \\
ElevenLabs & 14--16 & 23--28 & 27 \\
PyannoteAI & 75 & 101 & 80 \\
AssemblyAI & 14 & 22 & 20 \\
Deepgram & 26 & 26 & 16 \\
OpenAI & 71 & 124 & 62 \\
SE-DiCoW & 24 & 32 & 34 \\
WhisperX & 20 & 22 & 23 \\
\bottomrule
\end{tabular}}
\tableinfo{Counts exclude explicit unknown-speaker labels.}
\endgroup
\end{minipage}

Table~\ref{tab:persistent-identity-counts} makes identity proliferation visible beside the word-weighted SI-cpWER result. ThyVoice remains close to the 12-speaker reference on Clean and creates substantially fewer identities than most systems under noise. Count alone cannot reveal merges, unknown-speaker duration, or how speech is distributed across identities; it is not a complete fragmentation analysis. Appendix D reports complementary local clustering diagnostics.

\subsubsection{Local speaker activity}\label{local-speaker-activity}

\begin{widetableblock}
\noindent\begin{minipage}{\linewidth}
\noindent\hspace*{0.5in}\begin{minipage}{\dimexpr\linewidth-1in\relax}
\captionof{table}{Local-speaker DER/JER by condition and output surface.}\label{tab:local-der-jer}
\end{minipage}
\par
\begingroup\footnotesize
\setlength{\tabcolsep}{10pt}
\noindent\makebox[\linewidth][c]{%
\begin{tabular}{@{}llrrrr@{}}
\toprule
Component & Surface & Clean $\downarrow$ & Noisy $\downarrow$ & CHiME-6 $\downarrow$ & Mean $\downarrow$ \\
\midrule
DiariZen md-v2 & native & 18.32 / 23.65 & 29.51 / 37.14 & 68.24 / 68.75 & 38.69 / 43.18 \\
DiariZen md & native & 19.55 / 25.10 & 29.82 / 37.64 & 68.99 / 69.19 & 39.45 / 43.98 \\
Precision-2 & native & 25.36 / 33.50 & 32.99 / 43.51 & 68.86 / 69.12 & 42.40 / 48.71 \\
Pyannote Community-1 & native & 30.36 / 38.42 & 37.50 / 48.08 & 72.35 / 73.81 & 46.74 / 53.44 \\
GPT-4o Transcribe Diarize & segments & 43.09 / 51.43 & 46.87 / 55.02 & 82.63 / 85.02 & 57.53 / 63.82 \\
Scribe v2 & word turns & 44.13 / 49.22 & 48.06 / 53.55 & 81.18 / 79.91 & 57.79 / 60.90 \\
Universal-3 Pro & word turns & 42.62 / 46.80 & 48.15 / 54.05 & 74.76 / 76.05 & 55.18 / 58.97 \\
Nova-3 & word turns & 41.97 / 48.07 & 59.78 / 68.14 & 92.55 / 92.44 & 64.77 / 69.55 \\
\bottomrule
\end{tabular}}
\endgroup
\par\vspace{3pt}
\noindent\hspace*{0.5in}\begin{minipage}{\dimexpr\linewidth-1in\relax}
\footnotesize\textit{Notes.} OpenAI CHiME-6 is aggregated over chunks because speaker labels reset between chunks.
\end{minipage}
\par\vspace{4pt}
\end{minipage}
\end{widetableblock}

Table~\ref{tab:local-der-jer} shows that ThyVoice's DiariZen md-v2 component has the lowest DER/JER on the evaluated output surfaces in every condition. This local-speaker surface gives ThyVoice a stronger base for evidence gating, but lexical and identity errors still constrain the final transcript.

Word-labelled outputs require consecutive same-speaker words to be reconstructed into speaker turns. We use one fixed 0.25 s maximum gap for ElevenLabs, AssemblyAI, and Deepgram; native timelines remain unchanged. This is an offline DER/JER reconstruction rule and does not alter the accepted cpWER or SI-cpWER runs. The parameter changes DER and JER without changing the provider response, so Appendix C reports the complete merge-gap sensitivity sweep.

\begin{widetableblock}
\noindent\begin{minipage}{\linewidth}
\noindent\hspace*{0.5in}\begin{minipage}{\dimexpr\linewidth-1in\relax}
\paragraph{Enrollment-relevant speaker evidence}\label{enrollment-relevant-speaker-evidence}\mbox{}\par
\captionof{table}{Enrollment-relevant diagnostics on NOTSOFAR; cells report Clean / Noisy.}\label{tab:enrollment-speaker-evidence}
\end{minipage}
\par
\begingroup\footnotesize
\noindent\makebox[\linewidth][c]{%
\begin{tabular}{@{}
  >{\raggedright\arraybackslash}p{(\linewidth - 12\tabcolsep) * \real{0.2150}}
  >{\raggedright\arraybackslash}p{(\linewidth - 12\tabcolsep) * \real{0.1000}}
  >{\raggedleft\arraybackslash}p{(\linewidth - 12\tabcolsep) * \real{0.1200}}
  >{\raggedleft\arraybackslash}p{(\linewidth - 12\tabcolsep) * \real{0.1300}}
  >{\raggedleft\arraybackslash}p{(\linewidth - 12\tabcolsep) * \real{0.1250}}
  >{\raggedleft\arraybackslash}p{(\linewidth - 12\tabcolsep) * \real{0.1300}}
  >{\raggedleft\arraybackslash}p{(\linewidth - 12\tabcolsep) * \real{0.1800}}@{}}
\toprule
\begin{minipage}[b]{\linewidth}\raggedright
Component
\end{minipage} & \begin{minipage}[b]{\linewidth}\raggedright
Surface
\end{minipage} & \begin{minipage}[b]{\linewidth}\raggedleft
Count MAE $\downarrow$
\end{minipage} & \begin{minipage}[b]{\linewidth}\raggedleft
Merge (\%) $\downarrow$
\end{minipage} & \begin{minipage}[b]{\linewidth}\raggedleft
Purity (\%) $\uparrow$
\end{minipage} & \begin{minipage}[b]{\linewidth}\raggedleft
Coverage (\%) $\uparrow$
\end{minipage} & \begin{minipage}[b]{\linewidth}\raggedleft
Overlap recall (\%) $\uparrow$
\end{minipage} \\
\midrule
DiariZen md-v2 & native & 0.36 / 0.50 & 1.7 / 11.2 & 96.2 / 91.2 & 95.2 / 92.0 & 78.5 / 53.0 \\
DiariZen md & native & 0.33 / 0.48 & 1.9 / 11.2 & 95.6 / 90.8 & 95.1 / 91.4 & 79.9 / 55.1 \\
Precision-2 & native & 0.81 / 1.12 & 4.1 / 11.2 & 93.9 / 87.5 & 95.0 / 93.7 & 68.0 / 53.6 \\
Pyannote Community-1 & native & 0.33 / 0.87 & 6.6 / 19.2 & 91.2 / 84.4 & 92.8 / 92.0 & 51.7 / 40.0 \\
GPT-4o Transcribe Diarize & segments & 1.43 / 2.90 & 11.6 / 10.0 & 84.2 / 81.2 & 93.8 / 93.0 & 0.1 / 0.2 \\
Scribe v2 & word turns & 0.49 / 0.62 & 10.9 / 15.5 & 92.7 / 89.7 & 84.4 / 81.5 & 0.0 / 0.0 \\
Universal-3 Pro & word turns & 0.32 / 0.82 & 6.1 / 14.5 & 94.1 / 87.9 & 83.4 / 81.2 & 0.0 / 0.0 \\
Nova-3 & word turns & 0.71 / 1.78 & 13.1 / 37.5 & 89.6 / 79.2 & 90.1 / 71.2 & 0.0 / 0.0 \\
\bottomrule
\end{tabular}}
\endgroup
\par\vspace{3pt}
\noindent\hspace*{0.5in}\begin{minipage}{\dimexpr\linewidth-1in\relax}
\footnotesize\textit{Notes.} For word-reconstructed outputs, 0.0 overlap recall means that the exposed surface emits no simultaneous-speaker activity. It does not describe the provider's internal model.
\end{minipage}
\par\vspace{4pt}
\end{minipage}
\end{widetableblock}

The diagnostics in Table~\ref{tab:enrollment-speaker-evidence} indicate risk before voiceprint enrollment. Purity must be read with coverage, while overlap recall shows whether mixed intervals are visible for rejection or repair. ThyVoice's DiariZen md-v2 component combines high purity and coverage with substantial overlap recall, although noise increases merging and hides more overlap. These indicators do not directly measure contamination inside a speaker embedding. Appendix D reports the full diagnostic matrix and CHiME-6 overlap detection.

\subsubsection{Lexical recognition}\label{lexical-recognition}

\noindent\begin{minipage}{\linewidth}
\begingroup\footnotesize
\setlength{\tabcolsep}{10pt}
\captionof{table}{Word-sequence error rate by condition.}\label{tab:lexical-wer}
\noindent\makebox[\linewidth][c]{%
\begin{tabular}{@{}llrrrr@{}}
\toprule
Text source & Evaluation mode & Clean $\downarrow$ & Noisy $\downarrow$ & CHiME-6 $\downarrow$ & Mean $\downarrow$ \\
\midrule
ElevenLabs Scribe v2 & diarized-STT output & 30.37 & 35.80 & 35.03 & 33.73 \\
AssemblyAI Universal-3 Pro & diarized-STT output & 31.94 & 39.69 & 39.88 & 37.17 \\
Deepgram Nova-3 & diarized-STT output & 31.68 & 50.31 & 82.60 & 54.86 \\
Parakeet TDT 0.6B v3 & mixed-audio recognizer & 32.69 & 42.26 & 48.43 & 41.12 \\
Qwen3-ASR-1.7B & mixed-audio recognizer & 33.11 & 43.44 & 52.27 & 42.94 \\
OpenAI GPT-4o Transcribe & standalone transcription API & 39.30 & 46.96 & 53.09 & 46.45 \\
WhisperX large-v2 & WhisperX recognizer output & 37.15 & 45.91 & 82.88 & 55.31 \\
\bottomrule
\end{tabular}}
\endgroup
\end{minipage}

Table~\ref{tab:lexical-wer} shows that ElevenLabs has the lowest WER on the evaluated emitted word sequences in every condition. This lexical result is consistent with ElevenLabs's cpWER lead, but that lead does not carry through to persistent attribution. Recognition errors carry into SI-cpWER, so improving the lexical layer can improve the final attributed score. The table scores the listed lexical surfaces, not end-to-end WER for the parent systems. Qwen3-ASR-1.7B is ThyVoice's recognizer and Parakeet TDT 0.6B v3 is the checkpoint used by PyannoteAI; both are evaluated here on the original mixed recording. Appendix E reports the full substitution, deletion, and insertion components.

\subsection{Overlap-policy ablation}\label{overlap-policy-ablation}

Table~\ref{tab:ablation-overlap} compares four overlap policies on the same 20 clean NOTSOFAR meetings, with pooled enrollment throughout. \textbf{Separation + rematching} recovers speaker-specific overlap audio and admits accepted streams to ASR and identity evidence. \textbf{Omit overlap} disables separation and excludes duration-eligible overlap regions from ASR and identity processing. Both \textbf{Mixture ASR} variants retain overlap in each active local speaker's ASR stream without separation: \textbf{clean ID} uses only non-overlapping audio for identity matching and enrollment, whereas \textbf{mixture ID} uses mixture-containing audio for those decisions.

\begingroup\small
\noindent\begin{minipage}{\linewidth}
\captionof{table}{Overlap policies on 20 clean NOTSOFAR meetings; pooled enrollment. Scores are percentages.}\label{tab:ablation-overlap}
\centering
\begin{tabular}{@{}lrr@{}}
\toprule
Overlap policy & cpWER $\downarrow$ & SI-cpWER $\downarrow$ \\
\midrule
Separation + rematching & \textbf{34.39} & \textbf{32.73} \\
Omit overlap & 55.39 & 54.12 \\
Mixture ASR + clean ID & 42.53 & 41.42 \\
Mixture ASR + mixture ID & 42.86 & 67.76 \\
\bottomrule
\end{tabular}
\end{minipage}
\endgroup

Separation and rematching achieve the lowest cpWER and SI-cpWER among the tested policies. Retaining overlap mixtures for ASR improves cpWER relative to omission when identity evidence remains non-overlapping. The two mixture policies have similar pre-identification cpWER, but mixture-containing identity evidence yields substantially higher SI-cpWER. These results distinguish overlap use for transcription from overlap use for persistent identity on this subset; they do not isolate a specific internal failure such as profile drift.

Additional enrollment ablations are reported in Appendix F.

%% file: sections/discussion.tex
Persistent attribution is not a restatement of transcript or per-recording quality. Lexical recognition and local attribution constrain the best result available to the identity layer, but the commercial ranking changes when one speaker map must hold across recordings. Identity resolution can add attribution error when local clusters are split or merged incorrectly, and it can repair local fragmentation by resolving several clusters to one person. SI-cpWER therefore measures a requirement that cpWER alone does not impose.

The local diagnostics show why the surface exposed to enrollment matters. Native timelines that expose simultaneous speaker activity allow an identity layer to reject or repair mixed evidence. A timeline that emits one speaker during true overlap cannot support that decision from timing alone. Word-labelled STT surfaces expose less information to a downstream identity layer, although this does not establish what the provider detected internally. The overlap-policy comparison supports treating transcription audio and identity evidence as separate selection decisions.

The error decompositions distinguish omission from incorrect emitted or attributed evidence within each layer. WER separates deletions from substitutions plus insertions; cpWER separates deletions from IAR; and DER separates missed speech from false alarm plus speaker confusion. At the persistent layer, the cpWER-to-SI-cpWER change and resolved identity count expose the effect of carrying identity across recordings. These profiles are compared within a layer, not added across metrics. Their downstream cost also differs: an omission reduces recall, while an incorrect speaker assignment can attach a fact, decision, or action item to the wrong person. Measuring that task-level harm remains future work.

%% file: sections/limitations.tex
\textbf{Longitudinal validity.} NOTSOFAR meetings are single sessions. We simulate persistence by processing recordings serially against one shared voiceprint store in a fixed order, rather than observing the same people recur across days. The results therefore measure identity consistency under this fixed sequence; they do not establish calibration over longer timescales or robustness to recording order.

\textbf{Diagnostic comparability.} The lexical and local-speaker diagnostics score the component or provider outputs available for each system, not one uniform interface. WER over speaker-labelled hosted output differs from recognizer-only WER, and native diarization timelines differ from turns reconstructed from word labels. Ordinary WER can also penalize valid alternative serializations of overlapped speech. We therefore use these values to diagnose exposed evidence surfaces and retain cpWER and SI-cpWER as the end-to-end attribution metrics.

\textbf{Dataset coverage.} The evaluation covers English single-channel audio. Noisy NOTSOFAR uses deterministic augmentation rather than deployed capture, and CHiME-6 contributes only two far-field evaluation sessions. The results therefore do not establish multilingual or code-switching performance or generalization across broader devices and acoustic environments.

%% file: sections/conclusion_future_work.tex
This paper evaluates speaker attribution beyond one recording. SI-cpWER uses one corpus-global speaker map and reveals ranking changes hidden by per-recording cpWER. ThyVoice leads commercial SI-cpWER in all three conditions and the full-panel mean.

Substantial errors remain across the evaluated systems; improving both transcription and persistent identity resolution is therefore important for systems that store or retrieve information by speaker. The layered diagnostics locate constraints in lexical recognition, local speaker assignment, and identity resolution. ThyVoice is designed around overlap-aware speaker evidence and gated identity updates. Future work should evaluate recurring speakers over real multi-day use and measure whether attribution errors change retrieved facts, summaries, decisions, or action-item ownership. Multilingual and code-switching evaluation should examine recognition, alignment, speaker matching, and enrollment in these settings.

%% file: sections/appendix.tex
\section{Run Provenance and Targeted Repeatability Checks}\label{appendix-a-run-provenance-and-targeted-repeatability-checks}

\textbf{Voiceprint-bank batching.} pyannote \texttt{/identify} accepts at most 50 voiceprints per request. When the stored bank exceeds that limit, we search it in 50-voiceprint batches, select the best above-threshold match across all batches, and enroll a new voiceprint only if no batch returns a qualifying match. The accepted full-corpus runs apply this rule whenever the bank exceeds 50 entries.

\textbf{Deepgram diarizer resolution.} Deepgram Nova-3 was requested with \texttt{diarize\_model=latest}. At the June 2026 evaluation date, Deepgram documented that alias as its v2 batch diarizer \cite{deepgramdocs}. We retain both the requested alias and its time-specific resolution because \texttt{latest} can advance.

We ran three full 129-meeting ElevenLabs draws for each NOTSOFAR condition. Repeated calls returned different local speaker counts in a minority of meetings, which can change the identities resolved by the shared voiceprint bank. The following table records every draw; the main results report the corresponding means and observed ranges.

\noindent\begin{minipage}{\linewidth}
\begingroup\small
\setlength{\tabcolsep}{6pt}
\captionof{table}{ElevenLabs repeatability across three runs per condition.}\label{tab:el-variance}
\noindent\makebox[\linewidth][c]{%
\begin{tabular}{@{}lrrrrrr@{}}
\toprule
Draw & cpWER $\downarrow$ & S/N $\downarrow$ & D/N $\downarrow$ & I/N $\downarrow$ & IAR $\downarrow$ & SI-cpWER $\downarrow$ \\
\midrule
clean 1 & 32.89 & 9.18 & 15.82 & 7.89 & 17.07 & 34.61 \\
clean 2 & 33.91 & 8.82 & 16.66 & 8.44 & 17.25 & 36.21 \\
clean 3 & 33.91 & 9.23 & 16.22 & 8.46 & 17.69 & 37.57 \\
clean mean (spread) & 33.57 (1.02) & 9.08 & 16.23 & 8.26 & 17.34 & 36.13 (2.96) \\
noisy 1 & 43.25 & 12.86 & 22.61 & 7.79 & 20.65 & 50.11 \\
noisy 2 & 43.66 & 13.02 & 22.75 & 7.90 & 20.92 & 54.65 \\
noisy 3 & 43.84 & 12.91 & 22.79 & 8.13 & 21.04 & 56.91 \\
noisy mean (spread) & 43.58 (0.59) & 12.93 & 22.72 & 7.94 & 20.87 & 53.89 (6.80) \\
\bottomrule
\end{tabular}}
\endgroup
\end{minipage}

Repeatability checks were targeted rather than panel-wide. ElevenLabs received three full 129-meeting draws per NOTSOFAR condition, AssemblyAI received repeated clean-subset checks, and the remaining condition cells are single accepted observations. Hosted endpoints differ in the randomness controls they expose \cite{pyannotestt, elevenlabsscribe, assemblyaidiarization, openaitranscribe}; no panel-wide variance ranking is claimed.

\clearpage

\section{cpWER Error Decomposition}\label{appendix-b-cpwer-error-decomposition}

The main cpWER table reports each condition and the condition mean. This appendix reports the complete per-condition edit decomposition. IAR is S/N + I/N, so cpWER = D/N + IAR.

\begingroup\small
\begin{longtable}[]{@{}lrrrrr@{}}
\caption{cpWER decomposition by condition.}\label{tab:cpwer-components}\tabularnewline
\toprule\noalign{}
System & cpWER $\downarrow$ & S/N $\downarrow$ & D/N $\downarrow$ & I/N $\downarrow$ & IAR $\downarrow$ \\
\midrule\noalign{}
\endfirsthead
\toprule\noalign{}
System & cpWER $\downarrow$ & S/N $\downarrow$ & D/N $\downarrow$ & I/N $\downarrow$ & IAR $\downarrow$ \\
\midrule\noalign{}
\endhead
\bottomrule\noalign{}
\endlastfoot
\textbf{Clean} & & & & & \\
ThyVoice & 35.53 & 7.80 & 25.39 & 2.34 & 10.14 \\
ElevenLabs & 33.57 & 9.08 & 16.23 & 8.26 & 17.34 \\
PyannoteAI & 39.51 & 7.04 & 28.75 & 3.72 & 10.76 \\
AssemblyAI & 37.24 & 6.87 & 27.47 & 2.90 & 9.77 \\
Deepgram & 55.14 & 11.37 & 32.53 & 11.24 & 22.61 \\
OpenAI & 47.22 & 12.97 & 23.87 & 10.38 & 23.35 \\
SE-DiCoW & 25.62 & 8.74 & 10.80 & 6.09 & 14.83 \\
WhisperX & 50.00 & 10.95 & 35.74 & 3.31 & 14.26 \\
\midrule\noalign{}
\textbf{Noisy} & & & & & \\
ThyVoice & 51.42 & 11.23 & 37.89 & 2.30 & 13.53 \\
ElevenLabs & 43.58 & 12.93 & 22.72 & 7.94 & 20.87 \\
PyannoteAI & 52.71 & 10.11 & 37.17 & 5.43 & 15.54 \\
AssemblyAI & 51.36 & 9.20 & 36.48 & 5.69 & 14.89 \\
Deepgram & 74.08 & 11.60 & 51.91 & 10.57 & 22.17 \\
OpenAI & 63.86 & 25.04 & 23.27 & 15.54 & 40.59 \\
SE-DiCoW & 43.12 & 14.68 & 19.72 & 8.72 & 23.40 \\
WhisperX & 60.29 & 13.51 & 42.54 & 4.23 & 17.74 \\
\midrule\noalign{}
\textbf{CHiME-6} & & & & & \\
ThyVoice & 53.76 & 13.45 & 37.79 & 2.52 & 15.97 \\
ElevenLabs & 44.25 & 15.28 & 19.90 & 9.07 & 24.35 \\
PyannoteAI & 56.43 & 12.35 & 39.44 & 4.64 & 16.99 \\
AssemblyAI & 63.77 & 12.39 & 38.78 & 12.60 & 24.99 \\
Deepgram & 88.87 & 7.79 & 78.37 & 2.71 & 10.50 \\
OpenAI & 71.45 & 15.70 & 50.68 & 5.07 & 20.77 \\
SE-DiCoW & 56.88 & 23.62 & 21.64 & 11.62 & 35.24 \\
WhisperX & 67.35 & 20.39 & 40.40 & 6.57 & 26.96 \\
\end{longtable}
\tableinfo{ElevenLabs Clean and Noisy values are means over three accepted draws.}
\endgroup

Deletion represents omitted reference content. IAR represents substituted or inserted content in the locally speaker-attributed transcript. Because cpWER is speaker-attributed, IAR can include lexical errors or real words assigned to the wrong local speaker; it is not a direct hallucination or truthfulness metric.

\section{Local Speaker Evaluation}\label{appendix-c-local-speaker-evaluation}

This appendix expands the local-speaker results summarized in the main paper and reports merge-gap sensitivity for the word-reconstructed surfaces. Component lineage is ThyVoice--DiariZen md-v2, SE-DiCoW--DiariZen md, and WhisperX--pyannote Community-1; hosted rows use the provider output surface. Native timelines are unchanged; word-labelled outputs use the fixed 0.25 s maximum same-speaker gap.

\subsection{DER/JER error decomposition}\label{derjer-error-decomposition}

\noindent\begin{minipage}{\linewidth}
\begingroup\footnotesize
\setlength{\tabcolsep}{6pt}
\captionof{table}{DER/JER decomposition by condition and output surface.}\label{tab:der-jer-components}
\noindent\makebox[\linewidth][c]{%
\begin{tabular}{@{}llrrrrr@{}}
\toprule
Component & Surface & DER $\downarrow$ & Miss $\downarrow$ & False alarm $\downarrow$ & Confusion $\downarrow$ & JER $\downarrow$ \\
\midrule
\multicolumn{7}{l}{\textbf{Clean}} \\
DiariZen md-v2 & native & 18.32 & 10.99 & 1.65 & 5.68 & 23.65 \\
DiariZen md & native & 19.55 & 12.27 & 1.46 & 5.82 & 25.10 \\
Precision-2 & native & 25.36 & 14.97 & 0.84 & 9.55 & 33.50 \\
Pyannote Community-1 & native & 30.36 & 19.84 & 1.01 & 9.51 & 38.42 \\
GPT-4o Transcribe Diarize & segments & 43.09 & 30.66 & 0.47 & 11.96 & 51.43 \\
Scribe v2 & word turns & 44.13 & 38.66 & 0.22 & 5.25 & 49.22 \\
Universal-3 Pro & word turns & 42.62 & 38.33 & 0.28 & 4.01 & 46.80 \\
Nova-3 & word turns & 41.97 & 34.09 & 0.36 & 7.53 & 48.07 \\
\midrule
\multicolumn{7}{l}{\textbf{Noisy}} \\
DiariZen md-v2 & native & 29.51 & 19.98 & 0.88 & 8.65 & 37.14 \\
DiariZen md & native & 29.82 & 20.15 & 0.91 & 8.76 & 37.64 \\
Precision-2 & native & 32.99 & 19.00 & 0.85 & 13.15 & 43.51 \\
Pyannote Community-1 & native & 37.50 & 23.20 & 1.19 & 13.11 & 48.08 \\
GPT-4o Transcribe Diarize & segments & 46.87 & 31.15 & 0.51 & 15.21 & 55.02 \\
Scribe v2 & word turns & 48.06 & 40.86 & 0.45 & 6.75 & 53.55 \\
Universal-3 Pro & word turns & 48.15 & 40.26 & 0.45 & 7.44 & 54.05 \\
Nova-3 & word turns & 59.78 & 48.74 & 0.30 & 10.74 & 68.14 \\
\midrule
\multicolumn{7}{l}{\textbf{CHiME-6}} \\
DiariZen md-v2 & native & 68.24 & 42.53 & 9.43 & 16.29 & 68.75 \\
DiariZen md & native & 68.99 & 42.40 & 9.72 & 16.87 & 69.19 \\
Precision-2 & native & 68.86 & 41.53 & 8.88 & 18.45 & 69.12 \\
Pyannote Community-1 & native & 72.35 & 44.20 & 9.00 & 19.14 & 73.81 \\
GPT-4o Transcribe Diarize & segments & 82.63 & 70.24 & 3.42 & 8.97 & 85.02 \\
Scribe v2 & word turns & 81.18 & 53.51 & 7.69 & 19.98 & 79.91 \\
Universal-3 Pro & word turns & 74.76 & 53.48 & 7.56 & 13.72 & 76.05 \\
Nova-3 & word turns & 92.55 & 86.64 & 1.96 & 3.95 & 92.44 \\
\bottomrule
\end{tabular}}
\endgroup
\end{minipage}

The high miss term on word-reconstructed surfaces is partly a property of the exposed output: silence between emitted words is not labelled as speaker activity, and simultaneous speaker activity is usually serialized. These results describe what a downstream application can recover from that surface, not the provider's internal diarizer independently of recognition.

\subsection{Word-to-turn merge-gap sensitivity}\label{word-to-turn-merge-gap-sensitivity}

Word-labelled APIs require a reconstruction rule before DER and JER can be computed. We fix 0.25 s for the standard comparison and then sweep 0, 0.25, 0.5, 1, 2, and 3 s offline on the same saved words and speaker labels. AssemblyAI has independently shown that this reconstruction parameter can materially change DER without changing the underlying words or speaker labels \cite{assemblyaidercpwer}. Their study uses a different collar, overlap policy, and dataset; we use it as methodological precedent, not as a numerical comparison.

\noindent\begin{minipage}{\linewidth}
\begingroup\small
\setlength{\tabcolsep}{5pt}
\captionof{table}{Merge-gap sensitivity of reconstructed-turn DER.}\label{tab:merge-gap-sensitivity}
\noindent\makebox[\linewidth][c]{%
\begin{tabular}{@{}lrrrr@{}}
\toprule
System & DER at fixed 0.25 s $\downarrow$ & Best observed DER $\downarrow$ & Best gap (s) & DER range over sweep \\
\midrule
\textbf{Clean} & & & & \\
Scribe v2 & 44.13 & 37.36 & 3 & 37.36--56.30 \\
Universal-3 Pro & 42.62 & 35.09 & 3 & 35.09--46.25 \\
Nova-3 & 41.97 & 38.16 & 3 & 38.16--43.95 \\
\midrule
\textbf{Noisy} & & & & \\
Scribe v2 & 48.06 & 40.39 & 3 & 40.39--58.87 \\
Universal-3 Pro & 48.15 & 41.64 & 3 & 41.64--50.57 \\
Nova-3 & 59.78 & 56.94 & 3 & 56.94--61.61 \\
\midrule
\textbf{CHiME-6} & & & & \\
Scribe v2 & 81.18 & 79.85 & 1 & 79.85--85.41 \\
Universal-3 Pro & 74.76 & 73.06 & 2 & 73.06--75.99 \\
Nova-3 & 92.55 & 92.46 & 2 & 92.46--92.77 \\
\bottomrule
\end{tabular}}
\endgroup
\end{minipage}

The sweep confirms that reconstructed-turn DER is partly a property of the post-processing rule. The direction and magnitude are provider- and condition-dependent, which is why the main comparison uses one disclosed gap rather than each provider's best setting.

\section{Detailed Local Speaker-Clustering Diagnostics}\label{appendix-d-detailed-local-speaker-clustering-diagnostics}

The diagnostics in this appendix explain local DER and JER. They do not use transcripts, persistent voiceprints, cpWER, or SI-cpWER. Each recording is partitioned at every reference and hypothesis boundary. Duration co-occurrence between reference speakers and predicted clusters then supports an optimal local mapping and the following derived quantities:

\begin{itemize}
\tightlist
\item
  \textbf{Count accuracy} is the percentage of recordings with the correct number of local speakers; count MAE is the mean absolute count error.
\item
  \textbf{Fragmentation} means one reference speaker has nonzero support from more than one predicted cluster. \textbf{Merging} means one predicted cluster has nonzero support from more than one reference speaker.
\item
  \textbf{Cluster purity} is the duration-weighted share of each predicted cluster assigned to its dominant reference speaker. Contamination is one minus purity.
\item
  \textbf{Coverage} is the share of single-speaker reference time on which the system emits at least one speaker label. It does not require the label to be correct.
\end{itemize}

Cluster relations use single-speaker reference regions. Overlap evidence is preserved and evaluated separately. All nonzero relations are retained.

\subsection{Counts, clustering, and speaker coverage}\label{counts-clustering-and-speaker-coverage}

\begin{widetableblock}
\noindent\begin{minipage}{\linewidth}
\begingroup\footnotesize
\setlength{\tabcolsep}{5pt}
\captionof{table}{Speaker-count, clustering, and coverage diagnostics on NOTSOFAR.}\label{tab:clustering-diagnostics-native}
\noindent\makebox[\linewidth][c]{%
\begin{tabular}{@{}llrrrrrr@{}}
\toprule
Component & Surface & Exact count (\%) $\uparrow$ & Count MAE $\downarrow$ & Fragmented (\%) $\downarrow$ & Merged (\%) $\downarrow$ & Purity (\%) $\uparrow$ & Coverage (\%) $\uparrow$ \\
\midrule
\multicolumn{8}{l}{\textbf{Clean}} \\
DiariZen md-v2 & native & 69.0 & 0.36 & 6.0 & 1.7 & 96.2 & 95.2 \\
DiariZen md & native & 70.5 & 0.33 & 5.3 & 1.9 & 95.6 & 95.1 \\
Precision-2 & native & 51.2 & 0.81 & 9.3 & 4.1 & 93.9 & 95.0 \\
Pyannote Community-1 & native & 69.8 & 0.33 & 2.8 & 6.6 & 91.2 & 92.8 \\
GPT-4o Transcribe Diarize & segments & 32.6 & 1.43 & 19.7 & 11.6 & 84.2 & 93.8 \\
Scribe v2 & word turns & 62.8 & 0.49 & 0.8 & 10.9 & 92.7 & 84.4 \\
Universal-3 Pro & word turns & 72.9 & 0.32 & 0.2 & 6.1 & 94.1 & 83.4 \\
Nova-3 & word turns & 52.7 & 0.71 & 2.7 & 13.1 & 89.6 & 90.1 \\
\midrule
\multicolumn{8}{l}{\textbf{Noisy}} \\
DiariZen md-v2 & native & 58.1 & 0.50 & 3.5 & 11.2 & 91.2 & 92.0 \\
DiariZen md & native & 58.9 & 0.48 & 3.0 & 11.2 & 90.8 & 91.4 \\
Precision-2 & native & 39.5 & 1.12 & 11.8 & 11.2 & 87.5 & 93.7 \\
Pyannote Community-1 & native & 38.8 & 0.87 & 0.7 & 19.2 & 84.4 & 92.0 \\
GPT-4o Transcribe Diarize & segments & 10.9 & 2.90 & 33.3 & 10.0 & 81.2 & 93.0 \\
Scribe v2 & word turns & 53.5 & 0.62 & 1.5 & 15.5 & 89.7 & 81.5 \\
Universal-3 Pro & word turns & 54.3 & 0.82 & 0.2 & 14.5 & 87.9 & 81.2 \\
Nova-3 & word turns & 21.7 & 1.78 & 3.5 & 37.5 & 79.2 & 71.2 \\
\bottomrule
\end{tabular}}
\endgroup
\end{minipage}
\end{widetableblock}

DiariZen md-v2 combines high coverage with the highest cluster purity among the native systems. Under noise, its purity falls by 5.0 points and its merged-cluster rate rises by 9.5 points. OpenAI retains more than 90\% single-speaker coverage but has much lower exact-count accuracy and purity, showing that emitting activity is not the same as assigning it consistently.

\subsection{Overlap detection}\label{overlap-detection}

This section evaluates whether native diarization timelines detect simultaneous speaker activity over the full corpus. This is an upstream identity-safety requirement: when a timeline emits only one speaker during true overlap, a downstream identity pipeline cannot identify and remove that mixed interval from timing alone. Overlap recall is therefore necessary, but not sufficient, for safe enrollment. Precision also matters because falsely marking clean speech as overlap can discard usable speaker evidence. NOTSOFAR reference overlap is transcript-aligned from word timings; CHiME-6 uses native activity annotations. A complete miss means that the system emits no speaker during reference overlap.

\noindent\begin{minipage}{\linewidth}
\begingroup\footnotesize
\setlength{\tabcolsep}{6pt}
\captionof{table}{Overlap detection by native output surface.}\label{tab:overlap-detection-full}
\noindent\makebox[\linewidth][c]{%
\begin{tabular}{@{}lrrrrr@{}}
\toprule
Component & Precision (\%) $\uparrow$ & Recall (\%) $\uparrow$ & F1 (\%) $\uparrow$ & Complete miss (\%) $\downarrow$ & Active-count MAE $\downarrow$ \\
\midrule
\textbf{Clean} & & & & & \\
DiariZen md-v2 & 93.6 & 78.5 & 85.4 & 0.9 & 0.17 \\
DiariZen md & 93.0 & 79.9 & 86.0 & 0.9 & 0.19 \\
Precision-2 & 95.3 & 68.0 & 79.4 & 1.2 & 0.21 \\
Pyannote Community-1 & 92.7 & 51.7 & 66.4 & 2.3 & 0.28 \\
GPT-4o Transcribe Diarize & 34.1 & 0.1 & 0.3 & 3.5 & 0.42 \\
\midrule\noalign{}
\textbf{Noisy} & & & & & \\
DiariZen md-v2 & 94.8 & 53.0 & 68.0 & 3.5 & 0.28 \\
DiariZen md & 94.2 & 55.1 & 69.5 & 3.8 & 0.29 \\
Precision-2 & 95.0 & 53.6 & 68.6 & 2.4 & 0.27 \\
Pyannote Community-1 & 92.0 & 40.0 & 55.8 & 4.2 & 0.33 \\
GPT-4o Transcribe Diarize & 31.3 & 0.2 & 0.4 & 3.9 & 0.43 \\
\midrule\noalign{}
\textbf{CHiME-6} & & & & & \\
DiariZen md-v2 & 65.0 & 24.5 & 35.6 & 21.4 & 0.65 \\
DiariZen md & 63.9 & 26.3 & 37.2 & 21.1 & 0.65 \\
Precision-2 & 70.2 & 24.8 & 36.7 & 21.2 & 0.63 \\
Pyannote Community-1 & 63.4 & 20.4 & 30.9 & 23.5 & 0.67 \\
GPT-4o Transcribe Diarize & 36.5 & 0.0 & 0.0 & 60.9 & 1.01 \\
\bottomrule
\end{tabular}}
\endgroup
\end{minipage}

Noise primarily reduces overlap recall while precision stays high for DiariZen and pyannote. CHiME-6 is harder for every native system. Word-reconstructed timelines are excluded because one-speaker-per-word output cannot faithfully represent simultaneous activity. For those providers, the exposed interface does not contain enough information to remove hidden overlap from timing alone; this is a limitation of the output surface, not a claim about the provider's internal model.

\section{Lexical Error Decomposition}\label{appendix-e-lexical-error-decomposition}

This appendix expands the WER results summarized in the main paper. The table scores emitted words without speaker labels. The hosted diarized-STT rows retain their emitted text; Parakeet and Qwen use mixed-audio recognizer output, OpenAI uses its standalone transcription API, and WhisperX uses its configured recognizer output.

\noindent\begin{minipage}{\linewidth}
\begingroup\footnotesize
\setlength{\tabcolsep}{5pt}
\captionof{table}{WER decomposition by condition.}\label{tab:lexical-wer-components}
\noindent\makebox[\linewidth][c]{%
\begin{tabular}{@{}lrrrr@{}}
\toprule
Text source & WER $\downarrow$ & S/N $\downarrow$ & D/N $\downarrow$ & I/N $\downarrow$ \\
\midrule
\textbf{Clean} & & & & \\
ElevenLabs Scribe v2 & 30.37 & 11.13 & 13.69 & 5.55 \\
AssemblyAI Universal-3 Pro & 31.94 & 4.88 & 25.82 & 1.24 \\
Deepgram Nova-3 & 31.68 & 7.16 & 22.89 & 1.63 \\
Parakeet TDT 0.6B v3 & 32.69 & 5.89 & 25.30 & 1.50 \\
Qwen3-ASR-1.7B & 33.11 & 6.32 & 23.83 & 2.96 \\
OpenAI GPT-4o Transcribe & 39.30 & 5.52 & 32.41 & 1.37 \\
WhisperX large-v2 & 37.15 & 4.78 & 31.25 & 1.13 \\
\midrule\noalign{}
\textbf{Noisy} & & & & \\
ElevenLabs Scribe v2 & 35.80 & 12.95 & 18.84 & 4.01 \\
AssemblyAI Universal-3 Pro & 39.69 & 6.75 & 31.83 & 1.10 \\
Deepgram Nova-3 & 50.31 & 7.38 & 41.74 & 1.19 \\
Parakeet TDT 0.6B v3 & 42.26 & 8.27 & 32.82 & 1.17 \\
Qwen3-ASR-1.7B & 43.44 & 8.32 & 32.51 & 2.61 \\
OpenAI GPT-4o Transcribe & 46.96 & 7.55 & 37.94 & 1.47 \\
WhisperX large-v2 & 45.91 & 7.24 & 37.60 & 1.07 \\
\midrule\noalign{}
\textbf{CHiME-6} & & & & \\
ElevenLabs Scribe v2 & 35.03 & 14.26 & 16.21 & 4.56 \\
AssemblyAI Universal-3 Pro & 39.88 & 9.14 & 28.94 & 1.80 \\
Deepgram Nova-3 & 82.60 & 2.87 & 79.28 & 0.45 \\
Parakeet TDT 0.6B v3 & 48.43 & 10.50 & 36.69 & 1.24 \\
Qwen3-ASR-1.7B & 52.27 & 9.29 & 40.98 & 2.00 \\
OpenAI GPT-4o Transcribe & 53.09 & 9.79 & 39.83 & 3.46 \\
WhisperX large-v2 & 82.88 & 2.56 & 80.00 & 0.32 \\
\bottomrule
\end{tabular}}
\endgroup
\end{minipage}

Deletion dominates most systems as acoustics become harder, especially Deepgram and WhisperX on CHiME-6. Qwen and Parakeet degrade less sharply, but neither matches ElevenLabs on this lexical surface.

\section{Additional Controlled Ablations}\label{appendix-f-additional-controlled-ablations}

Tables~\ref{tab:ablation-enroll} and~\ref{tab:sep} compare enrollment audio on the same 20-meeting NOTSOFAR subset in Clean and Noisy conditions. The ThyVoice configurations form a direct paired ablation; the hosted-provider comparisons are descriptive single-run contrasts.

\subsection{ThyVoice enrollment: pooled vs.~longest-turn}\label{thyvoice-enrollment-pooled-vs.-longest-turn}

\noindent\begin{minipage}{\linewidth}
\begingroup\small
\captionof{table}{ThyVoice enrollment ablation: SI-cpWER (\%) on 20 NOTSOFAR meetings per condition.}\label{tab:ablation-enroll}
\noindent\makebox[\linewidth][c]{%
\begin{tabular}{@{}lrr@{}}
\toprule
Config & Clean SI-cpWER $\downarrow$ & Noisy SI-cpWER $\downarrow$ \\
\midrule
pooled (default) & \textbf{32.8} & \textbf{48.3} \\
longest-turn & 33.3 & 50.7 \\
\bottomrule
\end{tabular}}
\endgroup
\end{minipage}

Pooling improves ThyVoice's SI-cpWER by approximately 0.5 percentage point on Clean and 2.4 points on Noisy (Table~\ref{tab:ablation-enroll}). This comparison does not isolate individual voiceprint gates.

\subsection{Enrollment pooling across comparison cascades}\label{enrollment-pooling-across-comparison-cascades}

For the comparison cascades, we compare voiceprints enrolled from the longest turn with voiceprints enrolled from all pooled speaker segments. Only the enrollment audio changes; matching and upsert stay fixed. This tests API-constrained audio pooling, not embedding pooling, because PyannoteAI exposes opaque voiceprints \cite{pyannotevoiceprint}. Pooled audio can amplify hidden mixed speech; ThyVoice pools only evidence that passes its overlap and eligibility gates.

\noindent\begin{minipage}{\linewidth}
\begingroup\small
\captionof{table}{Comparison-cascade enrollment ablation: SI-cpWER (\%) on 20 NOTSOFAR meetings per condition.}\label{tab:sep}
\noindent\makebox[\linewidth][c]{%
\begin{tabular}{@{}lrrr@{}}
\toprule
System & Longest turn $\downarrow$ & Pooled $\downarrow$ & \(\Delta\) \\
\midrule
\textbf{Clean} & & & \\
PyannoteAI & 49.30 & \textbf{39.87} & -9.43 \\
ElevenLabs & 44.11 & \textbf{36.49} & -7.62 \\
AssemblyAI & 43.40 & \textbf{38.47} & -4.93 \\
OpenAI & 62.92 & \textbf{49.17} & -13.75 \\
Deepgram & 72.21 & \textbf{61.39} & -10.82 \\
\midrule\noalign{}
\textbf{Noisy} & & & \\
PyannoteAI & \textbf{59.92} & 62.22 & +2.30 \\
ElevenLabs & 64.65 & \textbf{49.30} & -15.35 \\
AssemblyAI & 83.21 & \textbf{56.90} & -26.31 \\
OpenAI & 94.28 & \textbf{68.48} & -25.80 \\
Deepgram & 88.60 & \textbf{73.56} & -15.04 \\
\bottomrule
\end{tabular}}
\tableinfo{\(\Delta = \text{pooled} - \text{longest turn}\); negative values favor pooling. Bold marks the better result per row. Full-scale results use stitched-30s enrollment.}
\endgroup
\end{minipage}

Table~\ref{tab:sep} shows that pooling improves 9 of the 10 reported cells, but PyannoteAI reverses direction between Clean and Noisy. The ElevenLabs comparison also includes run-to-run variation. These are single, unpaired provider runs, so the sweep is descriptive rather than variance-controlled.